\PassOptionsToPackage{unicode}{hyperref}
\PassOptionsToPackage{hyphens}{url}
\documentclass[
]{article}
\usepackage{xcolor}
\usepackage{amsmath,amssymb}
\usepackage{iftex}
\ifPDFTeX
  \usepackage[T1]{fontenc}
  \usepackage[utf8]{inputenc}
  \usepackage{textcomp} 
\else 
  \usepackage{unicode-math} 
  \defaultfontfeatures{Scale=MatchLowercase}
  \defaultfontfeatures[\rmfamily]{Ligatures=TeX,Scale=1}
\fi
\usepackage{lmodern}
\ifPDFTeX\else
\fi
\IfFileExists{upquote.sty}{\usepackage{upquote}}{}
\IfFileExists{microtype.sty}{
  \usepackage[]{microtype}
  \UseMicrotypeSet[protrusion]{basicmath} 
}{}
\makeatletter
\@ifundefined{KOMAClassName}{
  \IfFileExists{parskip.sty}{%
    \usepackage{parskip}
  }{
    \setlength{\parindent}{0pt}
    \setlength{\parskip}{6pt plus 2pt minus 1pt}}
}{
  \KOMAoptions{parskip=half}}
\makeatother
\NewDocumentCommand\citeproctext{}{}

\makeatletter
 \let\@cite@ofmt\@firstofone
 \def\@biblabel#1{}
 \def\@cite#1#2{{#1\if@tempswa , #2\fi}}
\makeatother
\newlength{\cslhangindent}
\newlength{\csllabelwidth}
\newenvironment{CSLReferences}[2] 
 {\begin{list}{}{%
  \setlength{\itemindent}{0pt}
  \setlength{\leftmargin}{0pt}
  \setlength{\parsep}{0pt}
  \ifodd #1
   \setlength{\leftmargin}{\cslhangindent}
   \setlength{\itemindent}{-1\cslhangindent}
  \fi
  \setlength{\itemsep}{#2\baselineskip}}}
 {\end{list}}
\usepackage{calc}

\usepackage{bookmark}
\IfFileExists{xurl.sty}{\usepackage{xurl}}{} 
\hypersetup{
  pdftitle={huff: A Python package for Market Area Analysis},
  pdfauthor={ Freiburg, Germany ORCID: 0000-0001-5168-9846 EMail: geowieland@googlemail.com},
  hidelinks,
  pdfcreator={LaTeX via pandoc}}

\title{huff: A Python package for Market Area Analysis}
\author{\textbf{Thomas Wieland}\\
Freiburg, Germany\\
ORCID: 0000-0001-5168-9846\\
EMail: geowieland@googlemail.com}
\date{05 October 2026}

\begin{document}
\maketitle

\subsection{Summary}\label{summary}

Market area models, such as the \emph{Huff Model} and its extensions,
are used to estimate regional market shares and customer flows of retail
and service locations. In health geography, market area and
accessibility models are applied for the analysis of catchment areas and
spatial accessibility of healthcare locations. The \texttt{huff} Python
package provides a complete workflow for market area and spatial
accessibility analysis, including data import, construction of
origin-destination interaction matrices, basic model analysis, parameter
estimation from empirical data, calculation of distance or travel time
matrices, and map visualization. The package is modular and
object-oriented. It is intended for researchers in economic geography,
regional economics, marketing, geoinformation science, and health
geography. The software is openly available via the
\href{https://pypi.org/project/huff/}{Python Package Index (PyPI)}. Its
development and version history are managed in a public
\href{https://github.com/geowieland/huff_official}{GitHub Repository}
and archived at \href{https://doi.org/10.5281/zenodo.18639559}{Zenodo}.
A methodological handbook is also
\href{https://geowieland.github.io/huff_official/}{available on GitHub
Pages}.

\subsection{Statement of need}\label{statement-of-need}

\subsubsection{Research context}\label{research-context}

Market area models are used in economic geography, regional economics,
geoinformation science, and marketing, enabling the analysis and
forecasting of market areas and customer flows for retail and service
locations. The classical and most popular approach is the \emph{Huff
Model} (Huff 1962, 1963, 1964) and its numerous extensions, such as the
\emph{Multiplicative Competitive Interaction (MCI) Model} (Nakanishi and
Cooper 1974, 1982; Cooper and Nakanishi 1983). Typical research
applications include examining the influence of marketing or location
variables on consumer store choice, forecasting the revenue of new
locations, or predicting the impact of new locations on existing ones
(De Beule et al. 2014; Fittkau 2004; Li and Liu 2012; Mensing 2018; Oruc
and Tihi 2012; Suárez-Vega et al. 2015; Wieland 2015, 2018).

In health geography, such models are used to analyse catchment areas
with respect to medical practices and hospitals (Bai et al. 2023; Fülop
et al. 2011; Jia 2016; Latruwe et al. 2023; Rhein et al. 2025; Wieland
2018). In both cases, methods for calculating spatial accessibility are
used that are structurally related to market area models, such as the
\emph{Hansen Accessibility} (Hansen 1959) and the \emph{Two-Step
Floating Catchment Area (2SFCA) Analysis} (Luo and Wang 2003). The two
concepts are also increasingly being linked to methods for analyzing the
supply structure and accessibility of health locations (Liu et al. 2022;
Luo 2014; Rauch et al. 2023; Subal et al. 2021). These models are also
applied to other location-related contexts such as airports or
recreation facilities (Wang et al. 2022; Wang et al. 2026).

\subsubsection{Challenges in
application}\label{challenges-in-application}

There are several major challenges in model-based market area and
accessibility analyses:

\begin{itemize}
\item
  Model calibration based on observed data on consumer behavior and/or
  store sales is difficult because the models are nonlinear in their
  weighting parameters (Huff and McCallum 2008; Wieland 2017). For this
  purpose, the \emph{MCI Model} (Nakanishi and Cooper 1974, 1982; Cooper
  and Nakanishi 1983) has been developed as an econometric estimation
  technique. As this approach requires empirical market shares for
  fitting, it is applied in cases where customer-store interaction data
  was obtained by surveys or secondary data (Baviera-Puig et al. 2016;
  Latruwe et al. 2023; Oruc and Tihi 2012; Suárez-Vega et al. 2015;
  Wieland 2015, 2018). When only total sales of the locations
  investigated are available, nonlinear iterative fitting approaches may
  be applied (De Beule et al. 2014; Güßefeldt 2002; Haines Jr et al.
  1972; Li and Liu 2012; Liang et al. 2020; Mensing 2018; Orpana and
  Lampinen 2003; Wieland 2017). Due to the pronounced sensitivity of
  market area models to weighting schemes, the availability of multiple
  calibration approaches is essential in market area analysis.
\item
  Market area and accessibility models typically include travel times
  instead of Euclidean distances. Calculating travel times is based on
  graph theory network analysis and requires real street networks.
  Therefore, there is a need for GIS (Geographic Information System)
  support and/or access to an API providing calculations based on input
  origins and destinations (Huff and McCallum 2008; Miller and Shaw
  2015). It is extremely helpful for researchers if they can also
  complete this part of the market area analysis workflow within the
  analysis tool.
\item
  Researchers must choose and compare appropriate weighting functions
  and parameters, which may result in substantially different results.
  For input variables such as travel time, several weighting functions
  (e.g., power, exponential, logistic) are used, and the model results
  are compared using goodness-of-fit metrics (Bai et al. 2023; Latruwe
  et al. 2023; Li and Liu 2012; Orpana and Lampinen 2003). It is, thus,
  necessary that, within the market area analysis workflow, several
  weighting functions are available, and that researchers may compare
  different model specifications based on model evaluation metrics.
\end{itemize}

\subsubsection{\texorpdfstring{Functionality of the \texttt{huff}
package}{Functionality of the huff package}}\label{functionality-of-the-huff-package}

The \texttt{huff} package for Python v1.9.x essentially provides the
following features:

\begin{itemize}
\item
  \emph{Data management and preliminary analysis}: Users may load
  customer origins and supply locations from GeoDataFrames, point
  shapefiles, CSV, or XLSX files. Attributes of customer origins and
  supply locations (variables, weightings) may be set by the user. The
  next step is to create an \emph{interaction matrix} with a built-in
  function, on the basis of which all implemented models can then be
  calculated. Within an interaction matrix, \emph{transport costs}
  (distance or travel time between customer origins and supply
  locations) may be calculated with built-in methods.
\item
  \emph{Basic Huff Model analysis}: Given an interaction matrix, users
  may calculate probabilities and expected customer flows with respect
  to customer origins, and total market areas of supply locations.
\item
  \emph{Model fitting based on empirical data}: Given empirical data on
  customer flows, regional market shares, or total sales, users may
  calibrate market area models. Model parametrization may be conducted
  using the \emph{MCI Model}, by Maximum Likelihood optimization, or by
  a local optimization algorithm (Güßefeldt 2002, Wieland 2017).
  Additionally, machine learning-based market area models may be trained
  using empirical customer flow data. Evaluation metrics are available
  that are automatically calculated to assess the predictive fit of the
  models (e.g., \emph{R-squared}, \emph{RMSE}, \emph{MAPE}).
\item
  \emph{Accessibility analysis}: The package provides methods of
  accessibility analysis, which may be combined with market area
  analysis (\emph{Hansen Accessibility}, \emph{2SFCA Analysis}).
  Competitor accessibility may also be calculated directly in order to
  extend the \emph{Huff Model} in terms of the \emph{Competing
  Destinations Model} (Fotheringham 1985).
\item
  \emph{GIS tools}: The library includes auxiliary GIS functions for
  market area analysis (buffer, distance matrix, overlay statistics) and
  clients for OpenRouteService (Neis and Zipf 2008) and OpenStreetMap
  (Haklay 2008) for simple maps. All of them are implemented in the
  modeling functions but may also be used stand-alone.
\end{itemize}

\subsection{State of the field}\label{state-of-the-field}

To the best of our knowledge, no open-source Python package currently
provides market area analysis and parameter estimation for the Huff or
MCI Model. In particular, there is no known open-source software package
that covers the entire workflow of market area analyses, as described in
the ``Statement of need'' section. Some but not all of the
functionalities mentioned are implemented in R packages: Both the
\texttt{SpatialPosition} package (Giraud and Commenges 2025) and the
\texttt{huff-tools} package (Pavlis et al. 2014) provide basic Huff
Model analyses with two parameters, calculation of air distances, and
map visualization. The R package \texttt{MCI} (Wieland 2017) focuses on
model fitting based on empirical data, but does not provide processing
of geospatial data and the calculation of distances or travel times.
Accessibility analysis via 2SFCA Analysis is implemented in the R
package \texttt{accessibility} (Pereira and Herszenhut 2024). The
(almost) complete workflow for market area analyses using the Huff/MCI
Model is currently only implemented in proprietary GIS software, namely
the \emph{ArcGIS Business Analyst} by \emph{ESRI} (Esri 2025; Huff and
McCallum 2008).

\subsection{Software architecture and
documentation}\label{software-architecture-and-documentation}

\subsubsection{Design choices}\label{design-choices}

The \texttt{huff} package is organized into a modular architecture that
separates core modeling functionality from auxiliary helper modules. All
model-related classes, methods and functions are implemented in the
\texttt{models} module. Supporting functionalities are provided in
separate modules, organized thematically. For example, the \texttt{ors}
module provides an OpenRouteService client for retrieving travel time
matrices and isochrones, which may be directly accessed from the
\texttt{models} module. This design allows auxiliary functions to be
used independently of the core models (stand-alone). Configurations for
all models, classes, and functions are stored in a \texttt{config}
module.

The \texttt{huff} library follows an object-oriented design. The class
structure reflects the conceptual actors within a spatial market:
Customer demand locations are represented by the
\texttt{CustomerOrigins} class and supply locations by the
\texttt{SupplyLocations} class. Their connection is established via an
interaction matrix containing all possible origin-destination
combinations and the corresponding data, such as travel times and
location attributes. It is created from the location data using the
built-in function \texttt{create\_interaction\_matrix()} from the
\texttt{models} module, resulting in an instance of the
\texttt{InteractionMatrix} class. All implemented model analyses can be
performed from an \texttt{InteractionMatrix} object, with the individual
steps of the model calculations being methods of this class, e.g.,
\texttt{transport\_costs()} for adding distances or travel times,
\texttt{probabilities()}, \texttt{flows()}, and \texttt{marketareas()}
for Huff Model calculations, or \texttt{mci\_fit()} for a MCI Model
analysis. These model analyses return objects of specific classes for
each model, e.g., \texttt{HuffModel} and \texttt{MCIModel} for Huff and
MCI Models, respectively. This structure was chosen to ensure a
consistent workflow and a unified data structure, regardless of which
model analysis is to be performed. All mentioned classes include
\texttt{summary()} and \texttt{plot()} methods. Any object contains time
stamps for each change, which may be accessed with the
\texttt{show\_log()} method.

\subsubsection{Example workflow}\label{example-workflow}

A basic Huff Model analysis (without empirical parameter estimation) in
the \texttt{huff} package consists of the following steps: 

1. Load geospatial data of customer origins and supply locations 

2. Define their attributes and weightings 

3. Create an interaction matrix from origins and destinations, 
including the calculation of distances or travel times

4. Calculate utilities, regional market shares, and expected customer flows 

5. Calculate total market areas of all supply locations

\subsubsection{Documentation}\label{documentation}

The workflow above is illustrated in the \emph{Examples} section of the
package
\href{https://github.com/geowieland/huff_official/blob/main/README.md}{README}.
For advanced model analyses (e.g., model calibration) and other package
functionalities, the
\href{https://github.com/geowieland/huff_official/tree/main/examples}{examples}
folder in the \texttt{huff}
\href{http://www.github.com/geowieland/huff_official.git}{GitHub
repository} contains commented Python scripts. In addition to the API
documentation and examples, \texttt{huff} provides a methodological
handbook as a \href{https://geowieland.github.io/huff_official/}{GitHub
Pages site}. This manual documents the mathematical formulations and
theoretical foundations of the implemented models and provides guidance
on their application and interpretation. This combination of software,
examples, and methodological documentation is intended to support
reproducible and transparent application of the package in research.

\subsection{Research impact statement}\label{research-impact-statement}

The \texttt{huff} package has already been used in a health geography
project at Würzburg University Hospital to model the catchment areas for
pediatric oncology care; one publication citing the package has resulted
from this research project so far (Kapitza et al. 2026). The library is
actively used: since its first release in April 2025, it has been
downloaded 44,915 times from the Python Package Index (source:
\href{https://pepy.tech/project/huff}{pepy.tech}, accessed October 5,
2026).

\subsection{AI usage disclosure}\label{ai-usage-disclosure}

No AI tools were used for software design, implementation, or
decision-making. GitHub Copilot in Microsoft Visual Studio Code using
the GPT-5 mini model (by OpenAI) was used to assist in drafting and
refining docstrings for documentation. The corresponding guidelines and
constraints defined by the author are documented in
\texttt{AGENTS-docstrings.md} in the
\href{https://github.com/geowieland/huff_official}{public GitHub
repository}. The manuscript text was written without the use of AI
tools.

\subsection*{References}\label{references}
\addcontentsline{toc}{subsection}{References}

\protect\phantomsection\label{refs}

\end{document}